\documentclass[10pt,conference]{IEEEtran}
\usepackage{cite}
\usepackage{amsmath,amssymb,amsfonts}
\usepackage{graphicx}
\usepackage{textcomp}
\usepackage{xcolor}
\usepackage[hyphens]{url}
\usepackage{fancyhdr}
\usepackage{hyperref}
\usepackage{algorithmic}
\usepackage[cjk]{kotex}
\usepackage[linesnumbered, vlined, ruled, boxed]{algorithm2e}
\usepackage{subcaption}
\usepackage{multirow}
\usepackage{multicol}
\usepackage{makecell}
\usepackage{enumitem}
\usepackage{booktabs} 
\usepackage{array}
\usepackage{comment}
\newcommand{\thickhline}{\Xhline{3\arrayrulewidth}}
\usepackage{marginnote}
\usepackage{romannum}

\newcommand{\hpcayear}{2025}
\IEEEoverridecommandlockouts

\newcommand{\hpcasubmissionnumber}{1185}
\title{Ditto: Accelerating Diffusion Model via Temporal Value Similarity}

\def\hpcacameraready{} 

\newcommand\hpcaauthors{Sungbin Kim$^{\ast}$\thanks{$^{\ast}$Both authors contributed equally to this work.}, Hyunwuk Lee$^{\ast\dagger}$\thanks{$^{\dagger}$The author is working at Samsung Electronics now.}, Wonho Cho, Mincheol Park, and Won Woo Ro}
\newcommand\hpcaaffiliation{School of Electrical and Electronic Engineering, Yonsei University, Seoul, Republic of Korea}
\newcommand\hpcaemail{Emails: \{sungbin.kim, hyunwuk.lee, wonho.cho, mincheol.park, wro\}@yonsei.ac.kr}

\usepackage[nolist]{acronym} 
\begin{acronym}
\acro{DNN}{Deep Neural Network}
\acro{NPU}{Neural Processing Unit}
\acro{TPU}{Tensor Processing Unit}
\acro{NN}{Neural Network}
\acro{CNN}{Convolution Neural Network}
\acro{RNN}{Recurrent Neural Network}
\acro{LSTM}{Long Short-Term Memory}
\acro{GAN}{Generative Adversarial Network}
\acro{VAE}{Variational Autoencoders}
\acro{MAC}{Multiply and Accumulate}
\acro{FID}{Frechet Inception Distance}
\acro{IS}{Inception Score}
\acro{PE}{Processing Element}
\acro{DDIM}{Denoising Diffusion Implicit Model}
\acro{LDM}{Latent Diffusion Model}
\acro{SDM}{Stable Diffusion Model}
\acro{PTQ}{Post-Training Quantization}
\acro{LLM}{Large Language Model}
\acro{Defo}{Ditto Execution Flow Optimization}
\acro{BOPs}{Bit Operations}
\end{acronym}

\author{
  \ifdefined\hpcacameraready
    \IEEEauthorblockN{\hpcaauthors{}}
      \IEEEauthorblockA{
        \hpcaaffiliation{} \\
        \hpcaemail{}
      }
  \else
    \IEEEauthorblockN{\normalsize{HPCA \hpcayear{} Submission
      \textbf{\#\hpcasubmissionnumber{}}} \\
      \IEEEauthorblockA{
        Confidential Draft \\
        Do NOT Distribute!!
      }
    }
  \fi 
}

\fancypagestyle{camerareadyfirstpage}{%
  \fancyhead{}
  
  \fancyhead[C]{
    \ifdefined\aeopen
    \parbox[][12mm][t]{13.5cm}{\hpcayear{} IEEE International Symposium on High-Performance Computer Architecture (HPCA)}    
    \else
      \ifdefined\aereviewed
      \parbox[][12mm][t]{13.5cm}{\hpcayear{} IEEE International Symposium on High-Performance Computer Architecture (HPCA)}
      \else
      \ifdefined\aereproduced
      \parbox[][12mm][t]{13.5cm}{\hpcayear{} IEEE International Symposium on High-Performance Computer Architecture (HPCA)}
      \else
      \parbox[][0mm][t]{13.5cm}{\hpcayear{} IEEE International Symposium on High-Performance Computer Architecture (HPCA)}
    \fi 
    \fi 
    \fi 
    \ifdefined\aeopen 
      \includegraphics[width=12mm,height=12mm]{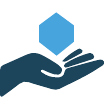}
    \fi 
    \ifdefined\aereviewed
      \includegraphics[width=12mm,height=12mm]{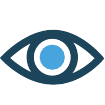}
    \fi 
    \ifdefined\aereproduced
      \includegraphics[width=12mm,height=12mm]{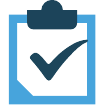}
    \fi
  }
  \fancyfoot[C]{}
}
\begin{document}
\maketitle

\ifdefined\hpcacameraready 
  \thispagestyle{camerareadyfirstpage}
  \pagestyle{empty}
\else
  \thispagestyle{plain}
  \pagestyle{plain}
\fi

\newcommand{\hpcaheight}{0mm}
\ifdefined\eaopen
\renewcommand{\hpcaheight}{12mm}
\fi

\begin{abstract}
Diffusion models achieve superior performance in image generation tasks. However, it incurs significant computation overheads due to its iterative structure.
To address these overheads, we analyze this iterative structure and observe that adjacent time steps in diffusion models exhibit high value similarity, leading to narrower differences between consecutive time steps.
We adapt these characteristics to a quantized diffusion model and reveal that the majority of these differences can be represented with reduced bit-width, and even zero.
Based on our observations, we propose the Ditto algorithm, a difference processing algorithm that leverages temporal similarity with quantization to enhance the efficiency of diffusion models.
By exploiting the narrower differences and the distributive property of layer operations, it performs full bit-width operations for the initial time step and processes subsequent steps with temporal differences. 
In addition, Ditto execution flow optimization is designed to mitigate the memory overhead of temporal difference processing, further boosting the efficiency of the Ditto algorithm.
We also design the Ditto hardware, a specialized hardware accelerator, fully exploiting the dynamic characteristics of the proposed algorithm.
As a result, the Ditto hardware achieves up to 1.5$\times$ speedup and 17.74\% energy saving compared to other accelerators. 
\end{abstract}
\section{Introduction}\label{sec:intro}
Diffusion models have demonstrated high performance in various image generation tasks such as image super-resolution~\cite{imagesuper1,imagesuper2}, video generation~\cite{video1,video2}, in-painting~\cite{inpaint1, inpaint2}, and text-to-image generation~\cite{texttoimage1, texttoimage2}.
Inspired by natural diffusion processes, it generates images through a reverse diffusion process that recursively denoises an image~\cite{diffusion1, diffusion2, ddpm, ddim}.
Through the process, it outperforms the previous image generation models (e.g., \ac{GAN}~\cite{gan, gan2} and \ac{VAE}~\cite{vae, vae2}) in terms of image quality and diversity~\cite{dm1,dm2,dm_survey}. 

Despite their advanced capabilities, diffusion models encounter significant computational demands~\cite{ddim, fastdiffusion}. 
Since the current time step of the reverse diffusion process requires the output of its former time step, the diffusion model cannot parallelize the execution of their time step~\cite{truncated,noise}.
Moreover, diffusion models employ a denoising model~\cite{bksdm, ldm} for each time step, which requires significantly increased computation compared to previous \ac{DNN} models (e.g., \ac{RNN}~\cite{rnn1, rnn2}, and \ac{LSTM}~\cite{lstm1, lstm2}) that adopt recurrent structure. 
These characteristics lead the diffusion models to be compute intensive~\cite{roofline} and have long execution time compared to other image generation models.

Due to the computational demand, quantization has emerged as a promising technique for diffusion models~\cite{tdq,ptqd,ptq4dm,qdiff}.
Previous software works~\cite{ptq4dm, qdiff} revealed that the activation value range gradually changes across time steps, caused by the inherent iterative feature of diffusion models, posing a significant challenge. 
This characteristic makes static quantization~\cite{olive, mokey} ineffective, as it leads to discrepancies between predefined scaling factors and actual value ranges.
To address the issue, previous works~\cite{ptq4dm, qdiff} utilized time step clustering technique based on the value range to determine more accurate scaling factors.

However, we consider dynamic change in activation values not as a challenge, but as an opportunity.
We assume that there is a potential value similarity between adjacent time steps due to the gradual changes on value range of activations in the diffusion models.
To verify our assumption, we analyzed the temporal value similarity of the reverse process. 
In our analysis, the data elements between adjacent time steps exhibit a high value similarity of 0.98.
Moreover, the temporal similarity is 0.67 higher than the spatial similarity inside activations, that is widely explored in vision-based neural network applications~\cite{diffy}.
Furthermore, this similarity results in a narrower value range for differences between adjacent time steps compared to the original activations.
Our experiments show that these temporal differences have a value range narrower up to 8.96$\times$ than the original activations.

To maximize the performance of diffusion models, we analyze the impact of the narrower value range of temporal differences in quantized models. 
Our analysis reveals that 96.01\% of the temporal differences between adjacent time steps require half bit-width for representation, with only 3.99\% of them necessitating full bit-width.
Moreover, zero temporal differences, indicating no change between time steps, account for 44.48\% of the total data elements.
These results demonstrate that the small temporal value differences between adjacent time steps enable the majority of them to be represented with reduced bit-width and even zero in quantized diffusion models.
In our analysis, leveraging both reduced bit-width and zero in the temporal differences would achieve 53.3\% \ac{BOPs}~\cite{bops,qdiff} reduction, while the state-of-the-art difference processing approaches~\cite{diffy} leveraging the spatial similarity achieves 38.8\%  \ac{BOPs} reduction.

Based on our observations, we propose the Ditto algorithm, a temporal difference processing approach that exploits temporal value similarity for efficient image generation in diffusion models.
The algorithm leverages the narrow value range of differences between adjacent time steps.
It executes the first time step with original activations and executes linear layers of further time steps with temporal differences.
It calculates the differences between adjacent time steps and executes linear layers only with these differences, using lower bit-width and utilizing zero skipping.
By leveraging the low computational intensity of the temporal differences, the algorithm effectively reduces computational overheads in diffusion models.

Additionally, we design Ditto Execution Flow Optimization (Defo) to dynamically optimize the execution flow of diffusion models using the Ditto algorithm.
Defo statically analyzes the dependencies of layers and checks for non-linear functions to reduce the memory overhead of loading the input and output of previous time steps when using temporal difference processing in layers.
Additionally, at run-time, it automatically determines the optimal execution flow (whether to execute with the original activations or differences) for subsequent time steps at second time step using the execution information from the first and second time steps of each layer.
Through the optimization, the Ditto algorithm achieves its effectiveness regardless of the type of diffusion model.

\begin{figure}[t]
    \centering
    \includegraphics[width=1\linewidth]{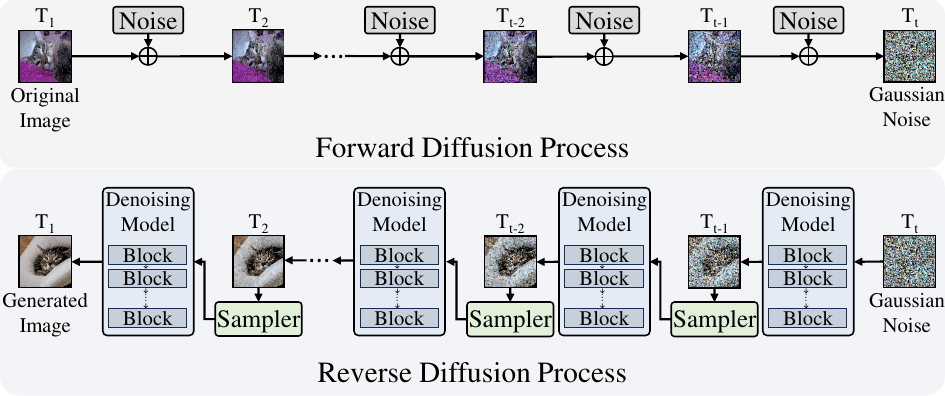}
    \caption{Image generation process of diffusion models. Diffusion models generate images through an iterative reverse process utilizing a denoising model. The reverse diffusion process starts from time step $T_{t}$ and ends at $T_{1}$.}
    \label{fig:diffusion_model}
\end{figure}

To exploit the benefits of the algorithm, we also design the Ditto hardware, a specialized hardware accelerator that supports dynamic sparsity and bit-width within the difference processing approach.
The hardware adopts adder tree-based \ac{PE} with corresponding encoder units to handle dynamic sparsity and bit-width simultaneously. 
It effectively calculates differences and supports dynamic sparsity within these differences through encoder units, and supports dynamic bit-width through \ac{PE}. 
Since a single PE design is utilized instead of an outlier PE to support mixed precision, the hardware fully accommodates the dynamic changes in throughput requirements for both lower and full bit-width operations introduced by the Ditto algorithm.
With this design, the proposed hardware effectively leverages the benefits of the Ditto algorithm, achieving high performance compared to other accelerators.

We summarize our contributions as follows:
\begin{itemize}[itemsep=0pt, leftmargin=*]
    \item We observe that the high similarity between adjacent time steps of the diffusion model results in a narrower value range of differences. Extending this into a quantized diffusion model and find out that 95.82\% can be represented with half bit-width including 44.76\% of zero values.
    \item Based on our observation, we propose the Ditto algorithm, a difference processing algorithm, that exploits value similarity in diffusion models with quantization to mitigate computational overheads of diffusion models. In addition, we design Defo, an optimization technique to maximize the performance across various diffusion models.
    \item We also design the Ditto hardware, a specialized hardware to fully support the dynamic characteristics of the algorithm.
    \item In our evaluation, the proposed hardware achieves up to 1.5$\times$ speedup and 17.74\% energy savings over the baseline.
\end{itemize}
\section{Diffusion Model}\label{sec:dm}

\begin{figure}[t]
    \centering
    \includegraphics[width=1\linewidth]{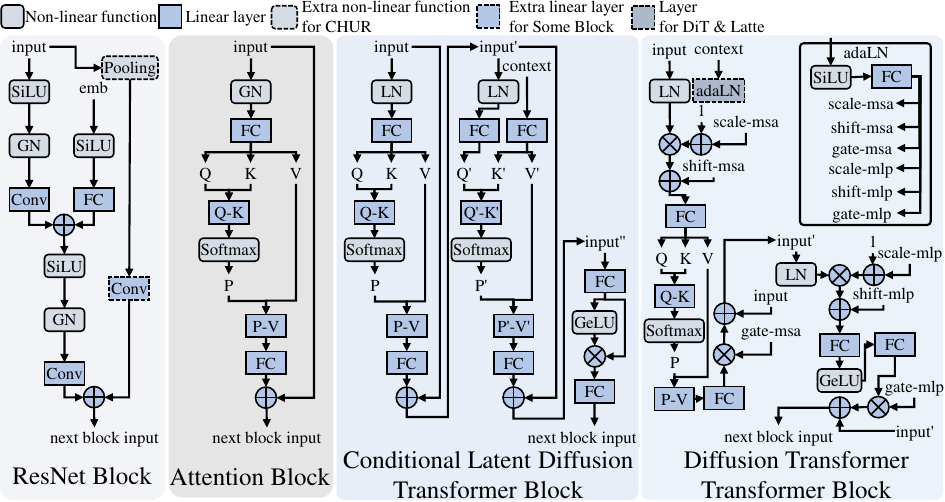}
    \caption{Various block structures of diffusion models in Table. \ref{tab:model}. GN and LN indicate group and layer normalization.}
    \label{fig:diffusion_layer}
\end{figure}

\subsection{Preliminaries of Diffusion Model}\label{subsec:background}
Diffusion models have recently achieved superior performance in various image generation tasks~\cite{diffusion1, diffusion2, diffusion3}. 
Inspired by the natural process of diffusion, diffusion models generate images by employing its reverse process~\cite{dm_survey}.
Fig. \ref{fig:diffusion_model} shows the image generation method in the diffusion model which comprises of forward and reverse diffusion process.
It first executes the forward diffusion process, which involves iteratively injecting noise into the original image.
Then, the reverse diffusion process generates an image by recurrently removing noise from the image.

Diffusion models utilize a neural network as a denoising model composed of sequentially connected blocks (group of layers) to reduce noise in reverse process~\cite{ddim,iddpm,ldm}.
Originally, the denoising model is composed of ResNet Blocks and Attention Blocks as shown in Fig. \ref{fig:diffusion_layer}. 
However, recently, various types of denoising models have been employed in diffusion models, each composed of different types of blocks~\cite{ldm,dit,latte}.
For instance, when using conditional techniques (IMG and SDM in Table. \ref{tab:model}), the Attention Block is replaced by a Conditional Latent Diffusion Transformer Block, resulting in a more complex structure. Moreover, diffusion transformers (DiT and Latte in Table. \ref{tab:model}) that use only transformer blocks without ResNet Blocks have also emerged~\cite{dit,latte}.
Since there is a wide variety of denoising models with different block structures, each model exhibits distinct layer dependencies and computation flow.
During the reverse process, the diffusion model uses the same network and weights for each time step, recursively feeding output from the previous time step ($T_{t}$) as the input for the current step ($T_{t-1}$)~\cite{ddim, iddpm, ddpm, ldm}.
Through the process, diffusion models achieve higher image quality and diversity than previous image generation \acp{DNN}, such as \ac{GAN}~\cite{gan, gan2} and \ac{VAE}\cite{vae, vae2}. 

Despite their advantages, diffusion models incur significant computational overheads in the reverse diffusion process due to their iterative characteristics and the high computational demands of the denoising model~\cite{bksdm, ddim, snapfusion,cal}.
Moreover, the recursive feedback mechanism prevents the parallelization of time steps, leading to long execution times~\cite{ddim, fastdiffusion} and high arithmetic intensity~\cite{roofline}.

\subsection{Value Similarity of Diffusion Model}\label{subsec:valsim}

\begin{figure}[!t]
    \centering
    \begin{subfigure}[t]{1\linewidth}
    \centering
    \includegraphics[width=1\linewidth]{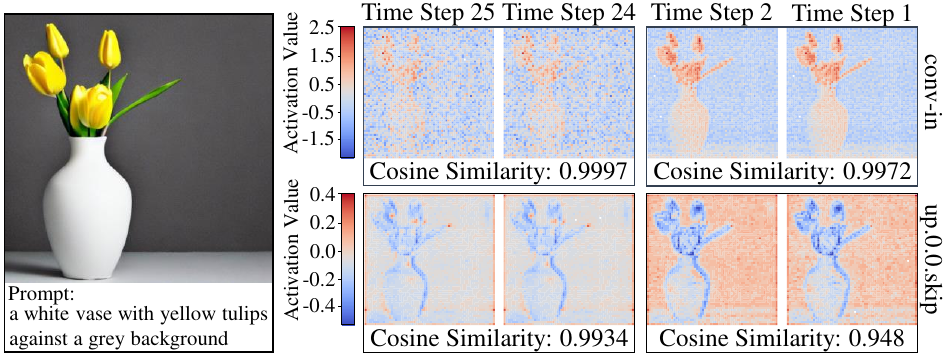} 
    \subcaption{Value heatmap and cosine similarity of activations between adjacent time steps in two layers of SDM~\cite{ldm}.}
    \vspace{0.5em}
    \label{fig:similarity_picture}
\end{subfigure}
\begin{subfigure}[t]{1\linewidth}
\centering
    \includegraphics[width=1\linewidth]{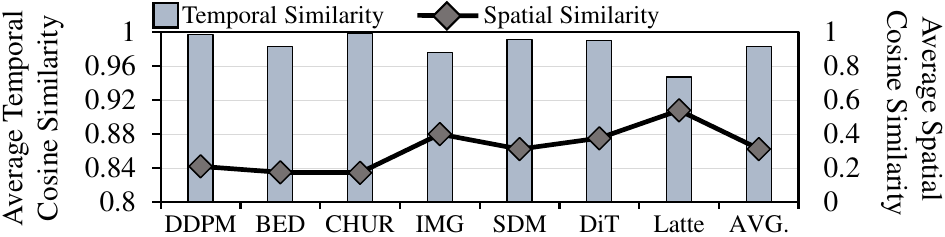} 
    \subcaption{
    Average temporal similarity of activations between adjacent time steps and average spatial similarity of activations across various diffusion models.}
\label{fig:similarity_graph}
\end{subfigure}
    \caption{
    Analyses on similarity of activations. 
    Cosine similarity is used for the similarity metric, ranging from -1 to 1, with the highest value as 1.}
    \label{fig:similarity}
\end{figure}



Since the diffusion models employ an identical denoising model and its weight for their entire time steps, a continuous adjustment to inputs would generate similarity of the data in layer operations. 
Thus, we assume that each data element within activations would exhibit a high degree of value similarity between consecutive time steps.
To validate our assumption, we conduct detailed analyses, focusing on the similarity between adjacent time steps. 
In Fig. \ref{fig:similarity}, the similarity within input activation of two layers (e.g., \textit{conv-in} and \textit{up.0.0.skip}) is measured through cosine similarity, which is widely used for measuring similarity between multi-dimensional data.
The analysis reveals that the similarity of activations exceeds 0.94 across these layers at various time steps (e.g., from time step 25 to 24, and 2 to 1).

We further measure the temporal similarity of all layers for every adjacent time step in various diffusion models, as shown in Fig. \ref{fig:similarity_graph}.
The details of the diffusion model benchmark in the analysis are provided in Table. \ref{tab:model}.
Our analyses demonstrate that the average cosine similarity in each model consistently surpasses 0.947, with an average similarity of 0.983 across various diffusion models.
We additionally measure the spatial similarity of layers, as previous research~\cite{diffy} leverages the spatial similarity inside the layer of computational imaging \acp{DNN}.
The results show that the diffusion models present a spatial similarity of 0.31 on average, which is lower than the temporal similarity.
Since the temporal similarity originates from the iterative process, the inherent characteristic of the diffusion models, the value similarity would exist in all diffusion models.

\section{Motivation}\label{sec:mot}
This section explores the design space of diffusion models associated with temporal value similarity.
We observe high temporal similarity results in low value differences between consecutive time steps.
Based on the observation, our analyses reveal that low temporal differences can effectively reduce bit-width in quantized diffusion models, potentially improving the performance of diffusion models.

\begin{figure}[!t]
\centering
\begin{subfigure}[t]{1\linewidth}
    \centering
    \includegraphics[width=1\linewidth]{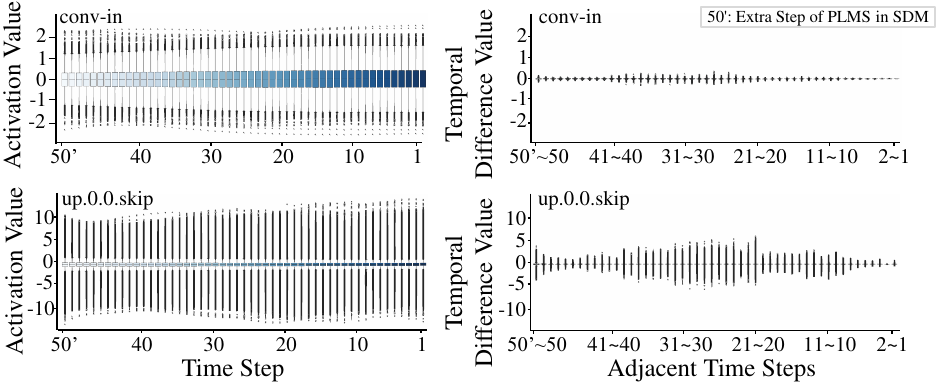} 
    \subcaption{Value range of activations and temporal differences across time steps in SDM~\cite{ldm}.}\vspace{0.5em}
    \label{fig:value_range_picture}
\end{subfigure}
\begin{subfigure}[t]{1\linewidth}
\centering
    \includegraphics[width=1\linewidth]{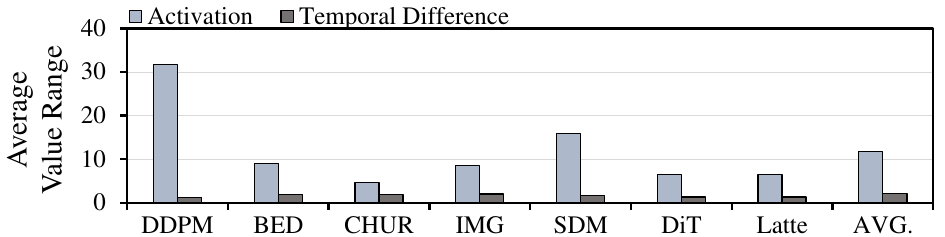} 
    \subcaption{Average value range of activations and temporal differences in various diffusion models.}
\label{fig:value_range_graph}
\end{subfigure}
\caption{Analyses on value range of activations and temporal differences in various diffusion models.} 
\label{fig:value_range}
\end{figure}

\subsection{Value Differences in Adjacent Time Step}
As high value similarity indicates a minimal difference in values, the range of the temporal differences tends to be narrower compared to the original activations.
Note that a reduced value range can improve computational efficiency~\cite{narrowvalue1, narrowvalue2, diffy}, we first conduct an experiment to examine the value ranges of both the original activations and the differences that can be obtained by subtracting each data element between consecutive time steps.
Fig. \ref{fig:value_range_picture} presents the experimental results on two layers (\textit{conv-in} and \textit{up.0.0.skip}) of the diffusion model.
In the \emph{conv-in} layer, our analysis reveals that the average value range of the original activations is 4.73, while the average range of the difference is merely 0.23.
Similarly, the original activations value range is 21.88, and the difference range is 4.83 on average in the \emph{up.0.0.skip} layer.

These narrower value ranges occur not only in specific time steps but across all time steps, showing the consistency of the narrower value range.
To verify these characteristics across various diffusion models, we conduct further experiments, comparing the average value range of the original activations and temporal differences in all layers, as shown in Fig. \ref{fig:value_range_graph}.
Using the same models as in Fig. \ref{fig:similarity_graph}, we calculate the average value range across all time steps.
The experiment results reveal that the value range of temporal differences presents an 8.96$\times$ narrower than the original activations on average.
Specifically, the value range of differences exposes up to 25.02$\times$ narrower in the DDPM, and at least 2.44$\times$ narrower in CHUR.
These results suggest that high temporal similarity reduces the range of differences between time steps, offering an opportunity to improve the computational performance of the diffusion models~\cite{narrowvalue2, diffy}.

\begin{figure}[!t]
    \centering
    \includegraphics[width=1\linewidth]{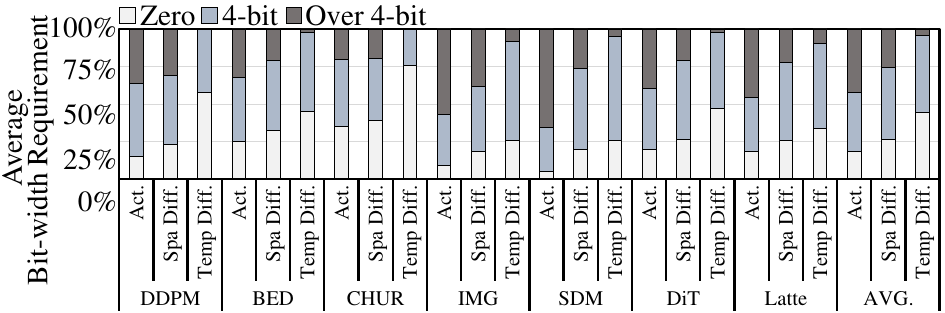} 
    \caption{
    Bit-width requirement of activations and differences in diffusion models. 
    Act., Spa Diff., and Temp Diff. denote original activations, spatial differences, and temporal differences.
    }
    \label{fig:motive_bit_req}
\end{figure}

\subsection{Advantages of Narrow Value Range}\label{sec:valsim_quant} 
Based on the above observation, we find out that a narrow value range in differences would be advantageous in the quantized diffusion model.
Since quantization compresses data into lower bit-width based on the value range of the data, a narrower value range in temporal differences could further reduce the bit-width required for operations~\cite{bitserial1, bitserial2, bitserial3}. 
To explore the potential benefit, we define the bit-width requirement as the minimum number of bits required to represent the value of the data.
With this term, we compare the bit-width requirement of differences of consecutive time steps with the original data in the quantized diffusion model.
Moreover, we also compare the bit-width requirement in the case of leveraging the spatial similarity inside layers.
For this case, the method of Diffy~\cite{diffy}, the state-of-the-art difference processing accelerator exploiting high spatial similarity in computational imaging DNNs, is adopted. 
Originally, Diffy targets only the spatial similarity of sliding convolution windows in convolution layers.
However, as the diffusion models consist of various types of layers~\cite{ldm,ddim,iddpm,dit,opensora}, we modify the Diffy method to support the similarity across the row dimension of input activation in fully connected layers and attention layers.

Fig. \ref{fig:motive_bit_req} shows our analysis results of the bit-width requirement in various quantized diffusion models.
In the analysis, the average bit-width requirement is measured for all data elements in diffusion models, quantized with simple dynamic quantization with 8-bit activation and weight.
The results show that zero temporal differences, indicating no change in values between time steps, constitute 44.48\% of the total temporal differences on average.
Since similar values are quantized into the same value~\cite{deepcompression}, our results indicate that most values between adjacent time steps are quantized to the same value owing to high temporal value similarity, resulting in a zero differences.
On the other hand, the original activations only exhibit zero value in quantization when the values are inherently zero or close to zero, thus, temporal differences show a 26.12\% higher ratio of zero than activations.
Moreover, due to the relatively low spatial similarity, a method leveraging temporal value similarity achieved an 18.04\% higher ratio of zero values compared to the spatial difference method.

We also find out that the values with lower bit-width requirements take a large portion of the temporal differences.
In the figure, the temporal differences that require a bit-width within 4 bits account for 51.52\%, even excluding zero-value differences.
Including zero temporal differences, those requiring 4-bit or fewer account for an average of 96.01\% of the total data elements in temporal differences.
These results indicate that only 3.99\% of the temporal differences require more than 4-bit for representation, which exhibits significant contrast to the original activations and spatial differences, where 42.28\% and 25.58\% require more than 4-bit.

Our analysis reveals that a significant portion of the temporal differences between consecutive time steps can be represented with reduced bit-width compared to the original activations in the quantized diffusion model.
Note that a lower bit-width reduces computational intensity~\cite{quanteff1, quanteff2}, reduced bit-width and a high portion of zero in differences can improve the computational efficiency of the diffusion model.

\begin{figure}[!t]
    \centering
    \begin{subfigure}[t]{1\linewidth}
    \centering
    \includegraphics[width=1\linewidth]{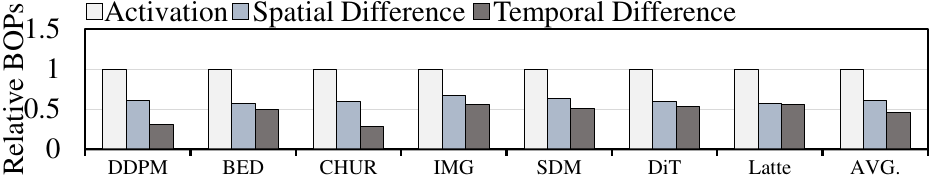} 
    \subcaption{
    Relative BOPs of the various methods across diffusion models.}
    \label{fig:potential_speedup}
\end{subfigure}
\begin{subfigure}[t]{1\linewidth}
\centering
    \includegraphics[width=1\linewidth]{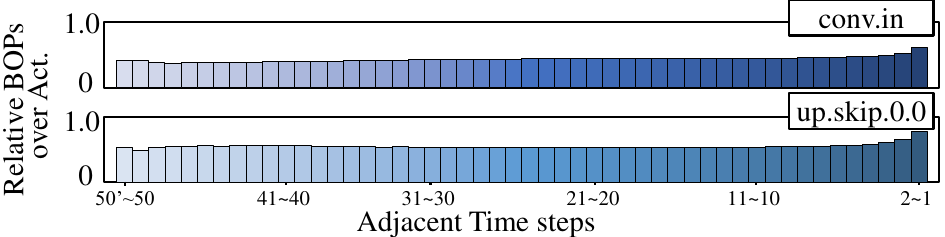}  
    \subcaption{
    Relative BOPs of the temporal difference approach compared to original activation across all adjacent time steps in SDM~\cite{ldm}.}
\label{fig:speedup_layer}
\end{subfigure}
    \caption{
    Analyses on BOPs in diffusion models.}
    \label{fig:potential}
\end{figure}

To verify our assumption, we analyze the relative number of \ac{BOPs}~\cite{bops, qdiff} for a single time step of diffusion models utilizing temporal differences compared to the original quantized model and the spatial difference method as shown in Fig. \ref{fig:potential_speedup}. 
The experiment utilizes our analysis results in Fig. \ref{fig:motive_bit_req}.
In the figure, the temporal difference approach can achieve 53.3\% and 23.1\% fewer \ac{BOPs} on average compared to the original models and the spatial difference method. 
Especially, DDPM and CHUR achieve 68.8\% and 71.5\% fewer \ac{BOPs} due to a higher proportion of zero temporal differences compared to other methods. 
These models exhibit 41.41\% and 35.53\% more zero values than the original activations and spatial difference method.  
We also examine whether this \ac{BOPs} reduction occurs at every time step. 
As shown in Fig. \ref{fig:speedup_layer}, the last few steps achieved a relatively lower \ac{BOPs} reduction because much denoising is required to generate the image in the final time steps.
However, even in these steps, a lower \ac{BOPs} is obtained compared to the original activations, and overall, consistent \ac{BOPs} reduction is achieved across most of the time steps. 
Consequently, the performance of the diffusion model can be boosted through reduced bit-width and zero skipping by utilizing temporal differences for all time steps.

\section{Ditto Algorithm}\label{sec:algorithm}

To exploit our observations, we propose the Ditto algorithm, a difference processing method leveraging temporal similarity in the diffusion model with execution flow optimization.
The algorithm consists of two techniques to apply difference processing in various types of layers in diffusion models.
The first part targets linear layers in the diffusion models, using the distributive property of linear algebra~\cite{linearalgebra1, linearalgebra2} to execute layer operations using temporal differences.
It takes advantage of the fact that the output of the linear layer at time step $t+1$ (i.e., previous time step) has already been computed, optimizing execution through reduced bit-width and zero-skipping.
To mitigate the potential overhead of temporal difference processing, we design the second technique, Ditto execution flow optimization (Defo).
With Defo, the proposed method automatically determines potential candidate layers that benefit from difference processing based on the layer information and adjusts the execution flow of each layer.

\subsection{Linear Layer Optimization}\label{sec:linear_ditto_algo}
\begin{figure}[t]
    \centering
    \includegraphics[width=1\linewidth]{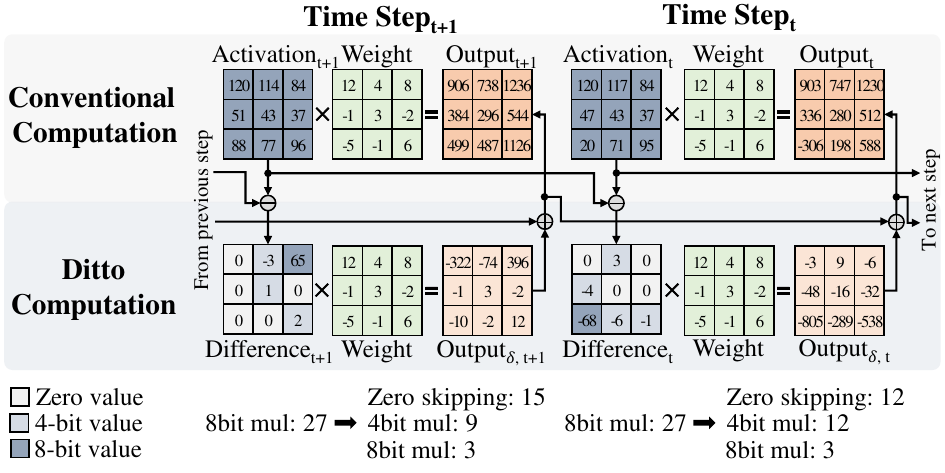}
    \caption{Process of linear layers in the Ditto algorithm.}
    \label{fig:delta_processing}
\end{figure}

\noindent\textbf{Convolution and Fully-connected Layers:} Fig. \ref{fig:delta_processing} presents an example of how the linear layer is executed in the Ditto algorithm.
To exploit the advantages of temporal differences, the Ditto algorithm executes layer operations with full bit-width for the first time step and then executes layer operations with differences between adjacent time steps.
For layer operations with temporal differences, the proposed algorithm comprises three stages.
In the first stage, the temporal differences between adjacent time steps are calculated by subtracting the input of the current time step from the input of the previous time steps.
Through the calculation, it can detect zero differences and the differences that can be represented in lower bit-width. 
After calculating the temporal differences, the Ditto algorithm executes the layer only with the differences in the second stage.
In this stage, the algorithm exploits reduced bit-widths and zero skipping for the layer operation, reducing the computational overheads of diffusion models.
In Fig. \ref{fig:delta_processing}, for example, it replaces twenty-seven 8-bit multiplication with nine 4-bit multiplication and three 8-bit multiplication at $Time \, Step_{t+1}$.
Finally, the Ditto algorithm applies summation between the result of difference processing and the previous time step output, as the third stage.
Since the diffusion models require numerous time steps to generate images~\cite{ddim, ddpm}, the proposed algorithm maximizes computational efficiency by utilizing three stages in whole time steps except the first time step.

\noindent\textbf{Attention Layers:} While convolution and fully-connected layers can be processed with the default difference processing algorithm, diffusion models also consist of attention layers.
Different from other linear layers, attention layers have operations (e.g., $Q \times K$ and $P \times V$) that multiply between two input matrices, changing across time steps.
If naively applying difference processing to attention layer, it requires three sub-operations for difference processing, $Q_{t+1} \Delta K$, $\Delta QK_{t+1}$, $\Delta Q \Delta K$, as $Q_tK_t=(Q_{t+1}+\Delta Q)(K_{t+1}+\Delta K)$ is equal to $Q_{t+1}K_{t+1}+Q_{t+1}\Delta K  +\Delta QK_{t+1} + \Delta Q \Delta K$.
However, since $Q_{t+1}\Delta K  +\Delta QK_{t+1} + \Delta Q \Delta K$ can be converted into $Q_t\Delta K  +\Delta QK_{t+1}$, the Ditto algorithm treat $Q_t$ and $K_{t+1}$ as weight and apply two sub-operations for attention layers. Also, the same mechanism applied to $P \times V$.
In our evaluation, the potential speedup of attention layers is always more than two of the original activations. 
Consequently, our optimization achieves higher performance than executing the attention layer with the original activations.

Moreover, we observe that in cross attention where \textit{context} is used as input, the values of the context remain unchanged across different time steps (second column in conditional latent diffusion model transformer block in Fig. \ref{fig:diffusion_layer}). 
Therefore, $K^{'}$ and $V^{'}$ do not change with varying time steps in the layer. 
With the observation, the Ditto algorithm treats $K^{'}$ and $V^{'}$ as weight in $Q^{'}\times K^{'}$ and $P^{'}\times V^{'}$, applying the same difference processing approach used in conventional linear layers.

\begin{figure}[t]
    \centering
    \includegraphics[width=1\linewidth]{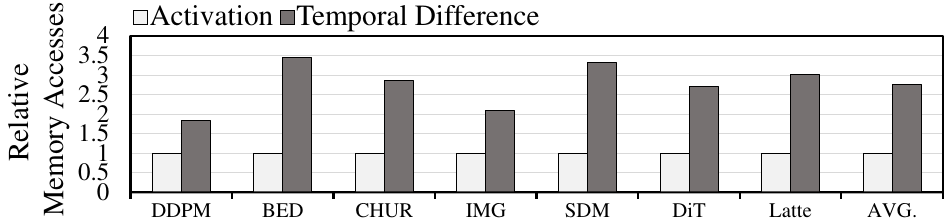}
    \caption{Relative memory accesses of the temporal difference processing across diffusion models.}
    \label{fig:mem_access}
    \vspace{-1em}
\end{figure}

\subsection{Execution Flow Optimization}\label{sec:dag_algo}
However, there are several challenges in applying the difference processing algorithm to the entire process of diffusion models.
First, since non-linear functions require original data to ensure numerical equivalence, the denoising model often needs the original data during execution.
Second, linear layer operations require additional memory accesses to obtain the linear layer input from the previous time step in order to calculate differences.
Therefore, some layers would be converted into memory-intensive operations due to the increased memory accesses and reduced computational intensity, even though diffusion models are compute-intensive networks~\cite{roofline}. 
In our analyses, temporal difference processing incurs 2.75$\times$ more memory accesses on average than original activation processing, as shown in Fig. \ref{fig:mem_access}.
Previous work, Cambricon-D~\cite{camd}, also addressed this issue by modifying non-linear functions such as SiLU and Group Normalization to reduce memory overhead. 
However, as shown in Fig. \ref{fig:diffusion_layer}, various diffusion models utilize a range of non-linear functions such as GeLU, Softmax, and Layer Normalization, limiting the effectiveness of their mechanism, particularly in models that do not use ResNet blocks, such as diffusion transformers.

\begin{figure}[t]
    \centering
    \includegraphics[width=1\linewidth]{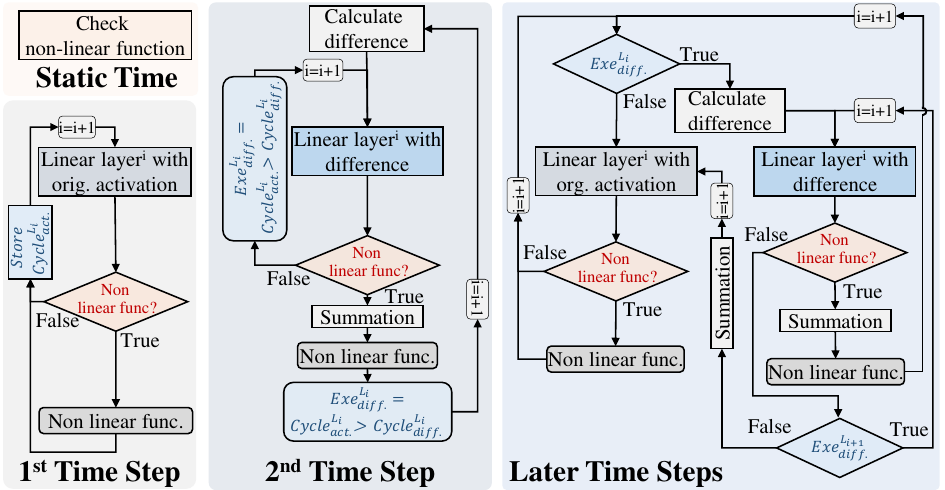}
    \caption{Overall flow of the Ditto algorithm.}
    \label{fig:dag}
\end{figure}

To mitigate memory overheads in various diffusion models, we propose the Ditto execution flow optimization (Defo).
It automatically determines whether to perform each linear layer operation using difference processing, adjusting the execution flow for layers.
Fig. \ref{fig:dag} shows a detailed execution process of the diffusion model with the difference processing and Defo.
In static time, Defo applies a computing graph analysis to find all non-linear functions and check the dependency of layers.
Based on the information, it modifies the difference processing algorithm by applying difference calculation and summation only before and after non-linear functions.

Even with the bypassing method, the issue of increased memory access may not be fully resolved. 
Therefore, it performs execution flow optimization at runtime to maximize the performance of the diffusion models.
Defo stores the cycle count of each layer during the first time step ($Cycle_{act.}$), which operates with the original activations. 
In the second time step, it dynamically determines the efficient execution type of each layer by comparing the cycle stored from the first step with the cycle using the difference processing algorithm ($Cycle_{diff.}$) which is determined by the number of zero and lower bit-width temporal differences. 
If $Cycle^{L_{i}}_{act.}$ is larger than $Cycle^{L_{i}}_{diff.}$, 
Defo set $Exe^{L_{i}}_{diff.}$ as True, enabling the layer to be executed using the difference processing method.

As we observed in Fig. \ref{fig:speedup_layer}, the ratio of BOPs reduction is consistent across time steps.
Based on the observation, Defo applies the execution flow of each layer determined in the second time step to all subsequent time steps.
In our evaluation, it successfully selects a more efficient execution flow with 92\% accuracy by using only first and second time step information, demonstrating the effectiveness and feasibility of our approach (see Fig. \ref{fig:eval_correct}).
For subsequent time steps, all layers operate with the execution type determined in the second time step. As in the second time step, difference calculation and summation are dynamically bypassed using layer dependency information, further reducing the memory overhead of the temporal difference processing method.

To further boost the performance of diffusion models, we introduce an additional optimization for layers that execute with the original activations.
In hardware utilizing temporal differences, they could leverage spatial differences with minor hardware modification.
Therefore, the Ditto algorithm is optimized to leverage spatial difference processing in the first time step and for layers determined by Defo to be executed with the original activations (defined as Defo$^{+}$).
Defo$^{+}$ calculates the spatial difference between the input data sequences and utilizes the difference in the same way as temporal difference processing.
As shown in Fig. \ref{fig:potential_speedup}, the spatial difference processing results in a reduction of \ac{BOPs} compared to original activation processing, while it is higher than the temporal difference processing.
Furthermore, since utilizing spatial difference processing calculates the difference within the intra-tensor, it does not require additional operations with the input and output of previous time steps, and thus does not incur memory access overhead.
Consequently, the computation reduction achieved through spatial differences enhances performance compared to using the original activations.

\section{Ditto Hardware}\label{sec:hardware}
While the Ditto algorithm enhances computational efficiency in diffusion models, its optimal performance would be constrained when implemented on general-purpose processors. 
The algorithm modifies the data in layer operation into a mix of zero, low, and full bit-width data.
Although a large portion of zero and low bit-width differences introduce the potential computational efficiency, it necessitates hardware architecture that supports both sparsity and mixed precision to fully exploit its potential.
A straightforward method to support the mixed precision is by incorporating outlier PEs.
However, the algorithm requires full bit-width data execution in the first time step and in layers determined by Defo to be executed with the original activations, making it difficult to design the optimal ratio between normal and outlier PEs.
To solve these design challenges, we propose the Ditto hardware, a specialized hardware accelerator to support both dynamic sparsity and bit-width in a single PE design.

\subsection{Hardware Overview}\label{subsec:hardware_overview}
\begin{figure}
    \centering
    \includegraphics[width=1\linewidth]{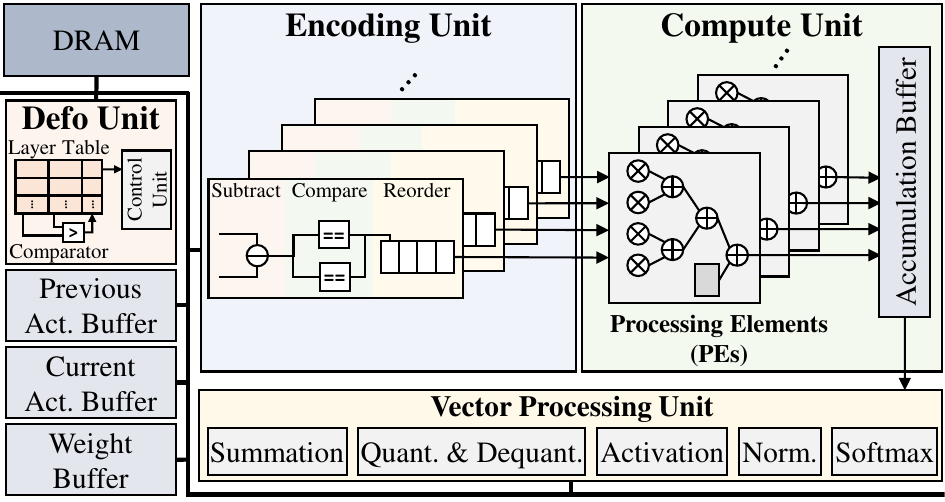}
    \caption{Overall architecture of the Ditto hardware.}
    \label{fig:delta_hardware_overall}
\end{figure}

Fig. \ref{fig:delta_hardware_overall} presents the overall architecture of the Ditto hardware.
It consists of four main components: Encoding Unit, Compute Unit, Vector Processing Unit, and Defo Unit.
Encoding Unit is a specialized hardware unit that calculates data differences and reorders the data elements to exploit the dynamic sparsity of Ditto.
It first computes the differences between consecutive steps, and classifies zero value, low bit-width, and full bit-width differences.
After classification, it reorders the data elements to skip zero value differences in Compute Unit and notates the full bit-width differences.

With reordered data, Compute Unit, a core unit of the proposed hardware, executes the actual layer operations of the diffusion model. 
We design Compute Unit as an adder tree based \ac{MAC} unit that supports two types of bit-width operation, a full bit-width (8-bit) and a low bit-width (4-bit) operation.
Since a large portion of the temporal differences is represented in 4-bit data, the bit-width of the baseline multiplier is set to multiplier between 4-bit activation and weights, supporting 8-bit activation by utilizing two multipliers and shifting logic. 
By leveraging reordered data by Encoding Unit, Compute Unit can skip zero and exploit the benefits of reduced bit-width while supporting full bit-width operation, ensuring numerical equivalent results with original operations. 
Vector Processing Unit operates the other special functions without linear layers such as non-linear functions. It also executes quantization and dequantization which is essential for quantized \ac{DNN}~\cite{deepcompression, mokey, accelerator,quantization1}. Also, it supports summation for the temporal difference processing.

The three components are designed as a pipelined architecture, a common technique for accelerators~\cite{pipeline1,pipeline2,pipeline3,mercury}, to maximize throughput and minimize latency.
We set the frequency of all components as the same, and the number of each component to support the maximum throughput of the Compute Unit executing in a low bit-width activation (4-bit).
Since the Ditto algorithm requires selective utilization of the three components, we design Defo Unit as a control unit, supporting the dynamic execution flow of the Defo.
It stores cycles of layers with original activations at the first time step and determines the type of execution for each layer.

\subsection{Detailed Hardware Design}\label{subsec:hardware_detail}

\begin{figure}[t]
    \centering
    \includegraphics[width=1\linewidth]{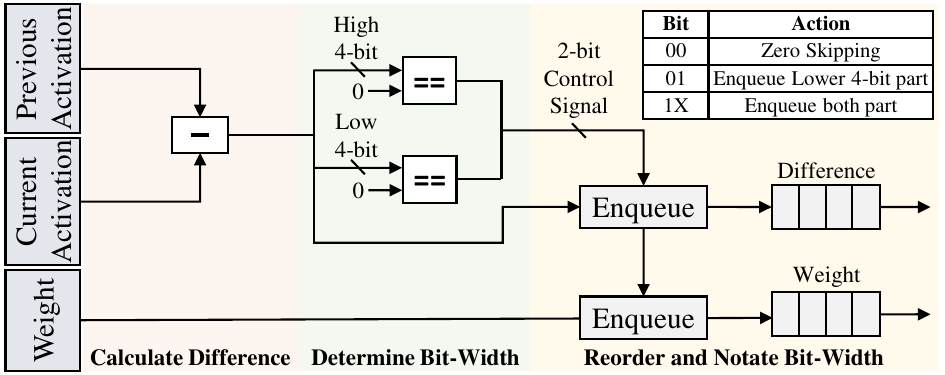}
    \caption{Detailed architecture of Encoding Unit.}
    \label{fig:delta_hardware_encoding}
\end{figure}

\noindent\textbf{Encoding Unit:} 
Fig. \ref{fig:delta_hardware_encoding} illustrates the detailed architecture of the Encoding Unit.
It has three main functionalities: calculating differences, classifying the bit-width requirement of data, and reordering data for zero skipping and notation of full bit-width data. 
To provide these functionalities, it consists of a subtractor, two comparators, and reordering logic.
Initially, it receives data elements from the previous and current time steps, calculating the difference between two data elements from these time steps using a subtractor.
Subsequently, it classifies each difference as a zero, low bit-width, and full bit-width. 
For the classification, it divides the differences into higher and lower bit parts and compares them with zero using two comparators.
The comparators identify zero value differences by detecting zero in both parts and low bit-width data by detecting zero only in the higher part. 
The rest of the comparison results are classified as full bit-width data. 
The outputs of these comparisons are synthesized into a 2-bit control signal utilized for the reordering process.
In the hardware that also exploits spatial differences, it only requires an offset register to store the spatial offset, and a multiplexer to switch the previous time step input into the spatial offset.

For the simple design of Compute Unit, we design Encoding Unit to eliminate zero differences and notate the full bit-width data during the reordering process.
This design allows Compute Unit to execute the reordered data elements without zero differences while also handling full bit-width data efficiently.
To support the functionality, Encoding Unit utilizes reordering logic to align data elements with the results of the classification
The reordering logic receives the 2-bit signal from the classification process and determines whether to skip the data element or enqueue it into a data queue.
By skipping the zero differences, Encoding Unit ensures the Compute Unit only executes data elements that require actual execution.

During the reordering process, Encoding Unit also reorganizes the data elements and notates those requiring full bit-width operations.
In the Compute Unit, a single multiplier is designed to multiply low bit-width data (i.e., 4-bit) with weight, while two multipliers are used for full bit-width data (i.e., 8-bit).
To support the design, Encoding Unit divides the data elements into high and low bit parts and enqueues them separately.
Then, it notates the data as full bit-width data to apply shift operation to the higher part.
Since Compute Unit supports shift operation per two multipliers for area efficiency, Encoding Unit reorders the data to align the high bit part with the multiplier with shift operations. 
As the order of accumulation is independent in the matrix multiplication, the high and low parts do not have to be accumulated in the first stage of the adder tree.
Therefore, it only needs to reorder the two parts of the data to ensure the higher part is directed to the multiplier with a shifter.
Consequently, it sends four 4-bit data with 2-bit flag for high bit part to a single adder tree unit of Compute Unit.
Note that Encoding Unit consists of simple logic, it can be implemented with low area overhead and high throughput.
Moreover, as subtraction and comparison can be combined into one cycle and queuing also can be done in one cycle, it can achieve low latency for its operation.

\begin{figure}
    \centering
    \includegraphics[width=1\linewidth]{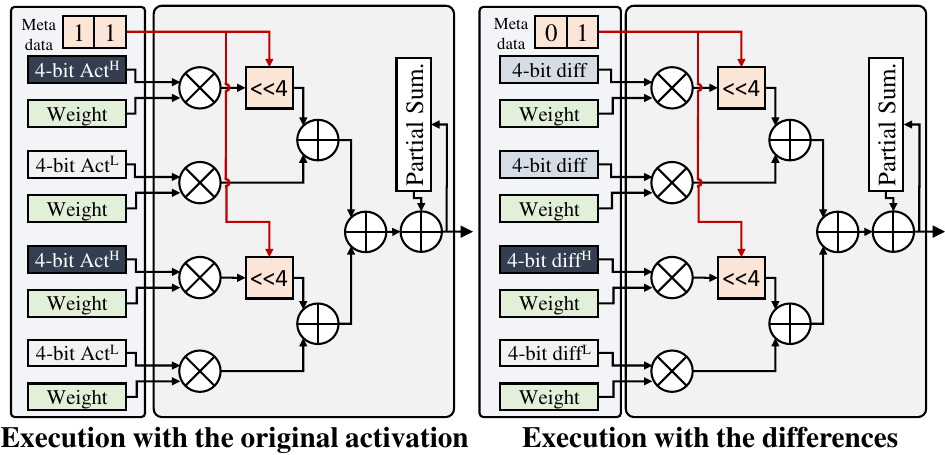}
    \caption{Detailed architecture and workflow of Compute Unit.}
    \label{fig:delta_hardware_compute}
\end{figure}

\noindent\textbf{Compute Unit:}
Fig. \ref{fig:delta_hardware_compute} illustrates the detailed architecture of Compute Unit. 
We design it as a set of adder tree based \ac{MAC} units that support two types of bit-width, 8-bit full bit-width and 4-bit low bit-width data.
Each \ac{PE} in the Compute Unit, consists of four multipliers that execute a multiplication between 4-bit data and weight, and a corresponding adder tree.
To support 8-bit operations, shifters are applied in the first adder stage.
As we design Encoding Unit to support reordering between the higher and lower part of the data, \ac{PE} only requires shifter logic per two multipliers for supporting 8-bit data.

As shown in the figure, the \ac{PE} receives the data and the metadata for higher part of full bit-width data from the Encoding Unit.
For 4-bit data, multiplication between activation and weight is straightforwardly executed through the multiplier, and then results are accumulated via an adder tree.
Besides, for 8-bit data, the \ac{PE} utilizes two multipliers to process each part of the data, the higher 4-bit and lower 4-bit parts, separately.
After multiplication, it recognizes the higher 4-bit through the metadata and applies shift operation to the result of a higher 4-bit part and then accumulates with an adder tree.
As previously mentioned, those two parts of the data do not have to be accumulated in the adder tree at once, they are accumulated through a partial sum register within \ac{PE}.

\noindent\textbf{Vector Processing Unit:} 
Vector Processing Unit supports various operations including activation functions, normalization, quantize and dequantize processes for quantized \ac{DNN}, and summation of results with previous outputs.
Since these operations are not required after every layer operation, Vector Processing Unit selectively executes them based on the model structure.
If layer operation does not need non linear function, it is bypassed to boost energy efficiency.

\noindent\textbf{Defo Unit:}
Defo Unit determines the execution type of layers.  
In the first time step, it executes the layer operation with full bit-width.
It also stores cycle information of each layer in a table to determine the execution flow of later time steps.
After the first time step, Defo Unit changes the execution type of all layers to temporal difference processing for the second time step.
Since the Defo optimizes the layer sequence in compile time, it dynamically skips Vector Processing Unit according to the layer sequence. 
During the second time step, Defo Unit monitors and stores the cycle of each layer in the table.
Then, it compares the second time step cycles with those from the first time step through compare logic and stores the results of the comparison in the table.
If the results indicate the effectiveness of the difference processing method, the Defo Unit maintains the execution type of the layer for all subsequent time steps accordingly. Otherwise, it changes the execution flow to the original activation execution and bypasses Encoder Units.

To support the functionality of the Defo Unit, the table requires entries for each layer. Based on our evaluation, the maximum number of layers of the diffusion model is 347, so we design it to have 512 entries to align a power of 2. 
Additionally, according to our evaluation, first time step and second time step cycle can be represented with 16-bit. 
Therefore, each entry is designed to have 33-bit, which includes 16-bit for the first time step cycle, 16-bit for the second time step cycle, and 1-bit for the later time step decision. 
Since the Defo unit is a control unit that only determines the type of execution path, it does not need to be scaled for throughput, unlike other hardware components. 
Therefore, it consumes only 0.01\% of the total hardware area.

\subsection{Operational Flow \& Communication}\label{subsec:operationflow}
To enable execution of the Ditto hardware, the CPU first sends the weights and input data to the DRAM in the Ditto hardware, along with the layer instructions to the Defo Unit. 
Since the DRAM and buffer memory store full bit-width (8-bit) data, we set the main interconnect between the DRAM, buffer memory, Encoder Unit, and Vector Processing Unit to operate on 8-bit data units.
Once all the data is stored, the Defo Unit initiates the operations of each unit based on the layer instructions.

For communication during layer operations, we design the interconnect connected to the Compute Unit to support different bit-width compared to the main interconnect.
The Encoder Unit sends difference and weight to the Compute Unit in sets of four 4-bit and four 8-bit data units, while the Compute Unit transfers accumulated data in 32-bit units to the Vector Processing Unit. 
These sets of data are aligned with the inputs and outputs of the processing units in the Compute Unit, ensuring that they remain unchanged whether the operation requires 4-bit or 8-bit data. 
After the Vector Processing Unit applies quantization to the output of the activation functions, it stores the resulting 8-bit data in the current activation buffer, and the hardware begins processing the next layer.

\section{evaluation}\label{sec:eval}
\subsection{Methodology \& Hardware Configuration}\label{subsec:hw_conf}
For the evaluation, we utilize seven diffusion models, shown in Table. \ref{tab:model}: one pixel-space unconditional diffusion model (DDPM), two latent-space unconditional diffusion models (BED, CHUR), two latent-space conditional diffusion models (IMG, SDM), and two diffusion transformer (DiT, Latte). 
We apply quantization to those models except for diffusion transformers with state-of-the-art quantization methods, Q-Diffusion~\cite{qdiff}. 
Since this method requires an offline calibration process to determine the scaling factor, we perform calibration on offline based on their repository.
In the similar way, dynamic quantization is adapted to two diffusion transformer models (DiT, Latte).

Table. \ref{tab:accuracy} presents our evaluation on the accuracy of the diffusion model with the Ditto algorithm. 
We employ various evaluation metrics commonly used in image generation tasks: \ac{FID}~\cite{fid}, \ac{IS}~\cite{is}, and CLIP Score (CS) \cite{clipscore}.
As shown in the table, the Ditto algorithm preserves the accuracy of all diffusion models compared to baseline FP32 models.

\begin{table}[t]
\caption{Evaluated Models, Datasets, and Samplers}
\centering
\resizebox{\linewidth}{!}
{
\begin{tabular}{>{\centering\arraybackslash}m{1.0cm}|>{\centering\arraybackslash}m{2.6cm}|>{\centering\arraybackslash}m{2.6cm}|>{\centering\arraybackslash}m{2.6cm}}
\thickhline
\textbf{Abbr.} & \textbf{Model} & \textbf{Dataset} & \textbf{Sampler \& Step}  \\
\hline
\hline
DDPM & DDPM~\cite{ddpm} & Cifar-10~\cite{cifar} & DDIM~\cite{ddim} 100 step \\
BED & Latent-Diffusion~\cite{ldm} & LSUN-Bed~\cite{lsun} & DDIM~\cite{ddim} 200 step\\
CHUR & Latent-Diffusion~\cite{ldm} & LSUN-Church~\cite{lsun} & DDIM~\cite{ddim} 200 step \\
IMG & Latent-Diffusion~\cite{ldm} & ImageNet~\cite{imagenet} & DDIM~\cite{ddim} 20 step\\
SDM & Stable-Diffusion~\cite{ldm} & COCO2017~\cite{coco} & PLMS~\cite{plm} 50 step \\ 
DiT & DiT-XL/2~\cite{dit} & ImageNet~\cite{imagenet} & DDIM~\cite{ddim} 250 step \\ 
Latte & Latte-XL/2~\cite{latte} & UCF-101~\cite{ucf101} & DDIM~\cite{ddim} 20 step \\ 
\thickhline
\end{tabular}
}
\label{tab:model}
\end{table}

\begin{table}[t]
\caption{Accuracy of Diffusion Models. FID is lower the better. IS and CS are higher the better.}
\centering
\resizebox{\linewidth}{!}{
\begin{tabular}{
   >{\centering\arraybackslash}m{1.2cm}|
   >{\centering\arraybackslash}m{1.6cm} 
  | >{\centering\arraybackslash}m{2.8cm} 
  | >{\centering\arraybackslash}m{2.8cm} 
   }
\thickhline
\textbf{Model} & \textbf{Metric} & \textbf{FP32} & \textbf{Ditto} \\ 
\hline
\hline
\multirow{1}{*}{\begin{tabular}[c]{@{}c@{}}DDPM \end{tabular}} & FID / IS & 4.143 / 9.084 & 4.406 / 9.288 \\ 
\hline
\multirow{1}{*}{\begin{tabular}[c]{@{}c@{}}BED \end{tabular}} & FID / IS & 2.962 / 2.227 & 5.897 / 2.338 \\
 \hline
 \multirow{1}{*}{\begin{tabular}[c]{@{}c@{}}CHUR \end{tabular}} & FID / IS & 4.100 / 2.715 & 3.743 / 2.714 \\
  \hline
  \multirow{1}{*}{\begin{tabular}[c]{@{}c@{}}IMG \end{tabular}} & FID / IS & 14.332 / 368.302 & 14.156 / 358.580 \\
  \hline
  \multirow{1}{*}{\begin{tabular}[c]{@{}c@{}}SDM \end{tabular}} & FID / IS / CS & 20.547 / 37.345 / 0.310 & 18.834 / 38.135 / 0.309 \\
 \hline
 \multirow{1}{*}{\begin{tabular}[c]{@{}c@{}}DiT \end{tabular}} & FID / IS & 18.659 / 482.372 & 17.178 / 475.694 \\
  \hline
   \multirow{1}{*}{\begin{tabular}[c]{@{}c@{}}Latte \end{tabular}} & IS & 70.589 & 71.254 \\
 \hline
 \thickhline
\end{tabular}
}
\label{tab:accuracy}
\end{table}

The Ditto hardware is evaluated using an open-source cycle-accurate simulator from Sparse-DySta~\cite{dysta}.
The simulator adopts hook functions provided by PyTorch~\cite{pytorch} to utilize dynamic sparsity in real \ac{DNN} workloads.
We modify this functionality to detect zero value, low bit-width, and full bit-width differences using actual input activation data in the diffusion model.
We design Tensor Core~\cite{tensorcore} as the baseline hardware in the simulator and extends to our hardware design. 
Additionally, it is modified to support spatial difference processing instead of the original activation execution and integrated into the simulator (defined as Ditto$^{+}$ hardware). 
To measure the area and power, Synopsys Design Compiler with a FreePDK 45nm library~\cite{pdk} is used to evaluate the core unit and CACTI~\cite{cacti} to measure the area and energy consumption of memory modules. 

For evaluating GPU system, we select an NVIDIA A100 GPU~\cite{a100}. 
We adopt the performance measurement method of the previous work~\cite{spatten} to compare GPU with proposed hardware.
Additionally, three baseline hardware accelerator designs are selected.
As a quantized model is utilized in our evaluation, an integer MAC unit-based Tensor Core-like unit is used as the baseline (defined as ITC). 
We additionally compare with hardware that employs difference processing methods: Diffy~\cite{diffy} and Cambricon-D~\cite{camd}.
Diffy utilizes spatial differences between sliding windows in convolution operations. 
Since the diffusion model also uses fully connected and attention layers, we extend the Diffy method to support difference processing along the row dimension of input activations in fully connected and attention layers for a fair comparison.
Cambricon-D exploits temporal differences in diffusion models. 
It also modifies the diffusion model using a software technique called sign-mask data flow, which bypasses difference calculation and summation for non-linear functions, such as SiLU and Group Normalization, to reduce the memory overhead of the temporal difference processing mechanism.
Unlike Ditto, it uses outlier PE designs to support full bit-width operations.
Also, it does not check layer dependency for non-linear functions and process attention layers with full bit-width operations. 
For a fair comparison, our dependency check technique and difference processing mechanism of attention layers integrate into the Cambricon-D.

\begin{table}[t]
\caption{
Hardware Configurations of Baseline and the Ditto Hardware
}
\centering
\resizebox{\linewidth}{!}
{
\begin{tabular}{>{\centering\arraybackslash}m{1.95cm}|>
{\centering\arraybackslash}m{1.6cm}|>
{\centering\arraybackslash}m{1.0cm}|>
{\centering\arraybackslash}m{0.8cm}|>
{\centering\arraybackslash}m{0.8cm}|>
{\centering\arraybackslash}m{0.7cm}|>
{\centering\arraybackslash}m{0.7cm}}
\thickhline
\textbf{Hardware} & \textbf{\# of PE} & \textbf{Bit-width of PE} & \textbf{Power (W)} & \textbf{SRAM (MB)} & \textbf{Area (mm\textsuperscript{2})} & \textbf{Freq.} \\
\hline
\hline
ITC~\cite{a100} & 27648 & A8W8 & 36.9 & \multirow{5}{*}{192} & \multirow{5}{*}{64.48} & \multirow{5}{*}{1GHz} \\
\cline{1-4}
Diffy~\cite{diffy} & 39398 & A4W8 & 33.6 & & & \\
\cline{1-4}
\multirow{2}{*}{Cambricon-D~\cite{camd}} & normal-38280  & A4W8 & \multirow{2}{*}{33.3} & & & \\
 & outlier-2552 & A8W8 & & & & \\
 \cline{1-4}
Ditto &39398 & A4W8 & 33.6 & & & \\
\thickhline
\end{tabular}
}
\label{tab:hw}
\end{table}

\begin{figure*}[t]
    \centering
    \includegraphics[width=1\linewidth]{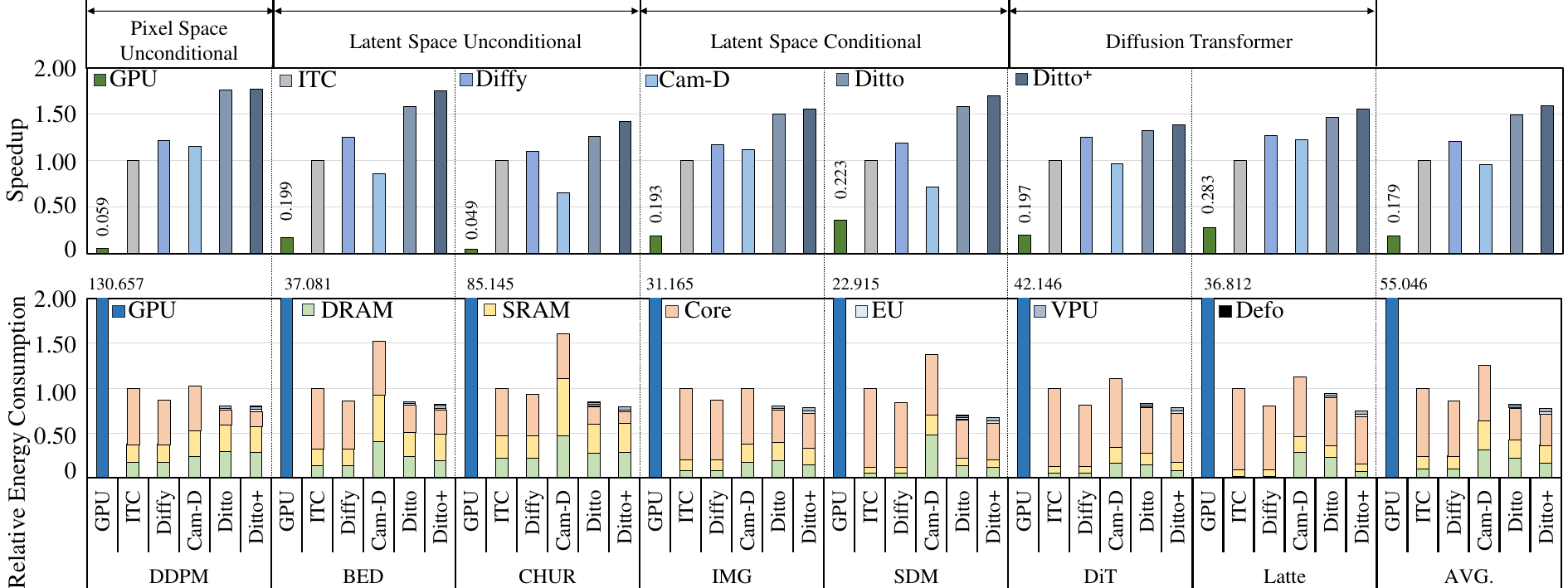}
    \caption{
   Comparison results of the various hardware in speedup (top), and relative energy consumption (bottom). Speedup and energy consumption are normalized to ITC.
   Cam-D, Ditto, Ditto$^{+}$, CU, EU, VPU, and Defo denote Cambricon-D, Ditto hardware, Ditto$^{+}$ hardware, Compute Unit, Encoding Unit, Vector Processing Unit, and Defo Unit, respectively.
    }
    \label{fig:Performance}
\end{figure*}

All baseline and Ditto hardware is designed to execute 8-bit activation and 8-bit weight (A8W8) quantized models, as this configuration preserves accuracy~\cite{qdiff, ptqd, ptq4dm}.
We adjust the number of \ac{PE} for iso-area comparison as shown in Table \ref{tab:hw}.
We fix the SRAM size and frequency to the same configuration across the hardware designs.
We also set the frequency of all components in the Ditto hardware (e.g., Encoding Unit, Vector Processing Unit, Defo Unit) to the same frequency as the \acp{PE}, as described in Section \ref{subsec:hardware_overview}.

\subsection{Performance Evaluation}\label{subsec:performance}
To evaluate the performance of Ditto hardware, we first compare it with other hardware designs in terms of speedup and energy consumption, as shown in Fig. \ref{fig:Performance}.
In the speedup evaluation, all hardware accelerator design achieves high performance over the GPU, due to the utilizing dedicated hardware design.
Compared to ITC, the Ditto hardware obtains 1.5$\times$ speedup on average, achieving the highest speedup across the other difference processing based hardware. Moreover, the Ditto$^{+}$ hardware exhibits 1.06$\times$ faster results compared to the Ditto hardware. This result aligns with the analysis in Section.  \ref{sec:valsim_quant}, spatial difference processing obtains potential speedup over the original activation execution. 
Diffy also exploits spatial differences. 
but shows 24\% lower performance compared to the Ditto hardware. 
Consequently, exploiting temporal difference processing is essential in diffusion models.

The Ditto hardware also shows a 1.56$\times$ speedup compared to Cambricon-D.
While Cambricon-D exploits temporal difference processing, their design requires full bit-width operation through outlier PEs. Consequently, with the same area budget, the Ditto hardware can accommodate more PEs to handle reduced bit-width operations and achieves additional speedup through zero skipping, resulting in better performance.
Additionally, while Cambricon-D mitigates memory overhead from temporal difference processing by utilizing sign-mask data flow for SiLU and Group Normalization, it cannot be applied to diffusion models using other non-linear functions such as GeLU, Softmax, and Layer Normalization (Fig. \ref{fig:diffusion_layer}).
However, our Ditto hardware can fully reduce memory overhead by automatically determining the optimal execution flow for each layer in all diffusion models through Defo.

\begin{figure}
    \centering
    \includegraphics[width=1\linewidth]{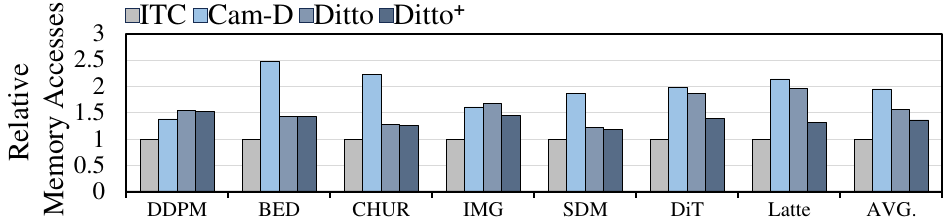}
    \caption{Relative memory accesses of the various hardware. Memory accesses are normalized to the ITC.}
    \label{fig:eval_memory_access}
\end{figure}

We also evaluate the energy consumption of the Ditto hardware with other hardware designs.
Similar to speedup experiments, all of the dedicated hardware achieves lower energy consumption over the GPU.
The Ditto and Ditto$^{+}$ hardware achieves 17.74\% and 22.92\% energy saving over ITC, which is larger than other difference processing based hardware.
These energy savings are due to reduced execution times, which lower the energy consumption of Compute Unit by exploiting both dynamic sparsity and bit-width.  
Diffy also shows similar results with speedup, achieving 14.3\% energy saving over ITC.
However, Cambricon-D exhibits higher energy consumption than ITC on average.
The result comes from a few benchmarks, such as BED, CHUR, and SDM, showing notably higher energy consumption.
In these benchmarks, significant memory overhead occurs by the temporal difference processing algorithm. This is due to the large input and output sizes of the linear layers that non-linear functions cannot be resolved by sign-mask data flow.
While the Ditto hardware also faces the overheads as shown in Fig. \ref{fig:eval_ditto}, it mitigates these overheads through Defo. As a result, the Ditto hardware achieves 43.24\% energy savings compared to Cambricon-D. 

Additionally, we evaluate the overhead of the additional components in the Ditto hardware.
Since we adopt a fully pipelined architecture in the accelerator design, the Ditto hardware overlaps the Encoding Unit, the Vector Processing Unit, and the Defo Unit with the execution of the Compute Unit.
As a result, the latency overheads of the Encoding Unit, the Vector Processing Unit, and the Defo Unit only account for 0.1\%, 0.17\%, and 0.1\% of the total latency, respectively.
The energy consumption of these units accounts for only 2.23\%, 2.9\%, and 0.0001\%, respectively in the Ditto hardware.

\begin{figure}
    \centering
    \includegraphics[width=1\linewidth]{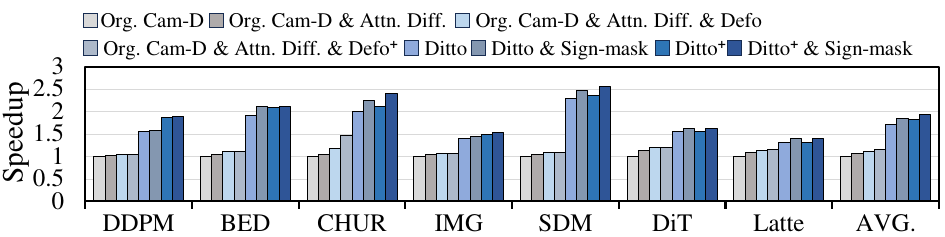}
    \caption{Comparison of Cambricon-D and the Ditto hardware with various software techniques in Cabmricon-D and Ditto. All designs utilize the layer dependency check technique.
    Speedup is normalized to the original Cambricon-D.}
    \label{fig:eval_camd}
\end{figure}

As temporal difference processing increases memory accesses, we evaluate memory access of hardware designs that leverage temporal differences.
In Fig. \ref{fig:eval_memory_access}, we find out that the memory accesses in all hardware designs are higher than the baseline ITC.
Cambricon-D incurs 1.95$\times$ more memory accesses than ITC, and Ditto and Ditto$^{+}$ show 1.56$\times$ and 1.36$\times$ more accesses, respectively.
The Defo algorithm automatically reduces memory overheads of memory-intensive layers, while Cambricon-D only reduces specific layers.
As a result, Ditto and Ditto$^{+}$ achieve fewer memory accesses than Cambricon-D and demonstrate greater generality.

In order to further demonstrate the advantages of our hardware over other designs utilizing temporal differences, we conducted a detailed comparison between the Ditto and Cambricon-D, which also leverages temporal differences for diffusion models.
Since the software optimization of Cambricon-D and Ditto can be applied to each other, we apply the software techniques of the Ditto algorithm to Cambricon-D and sign-mask data flow of Cambricon-D to the Ditto hardware, as shown in Fig. \ref{fig:eval_camd}.
In the figure, Cambricon-D achieves a 1.16$\times$ speedup when all Ditto algorithm techniques are applied and the Ditto and Ditto$^{+}$ hardware achieve 1.068$\times$ and 1.055$\times$ speedup through the sign-mask data flow.
These results indicate that both hardware can benefit from the software techniques of the others since those are complementary.

However, all of the Cambricon-D design shows lower performance than the Ditto hardware due to limitations in outlier PEs based design, performing original activation execution with a smaller number of PEs.
Consequently, even with the Defo technique, the cycles for original activation execution are too high, causing memory overhead reduction to be offset by compute overhead.
To effectively address memory overhead in temporal difference processing, a design like Ditto hardware, which dynamically selects bit-width, is more efficient.

\subsection{Design Space Exploration}\label{subsec:dsp}
\begin{figure}[t]
    \centering
    \includegraphics[width=1\linewidth]{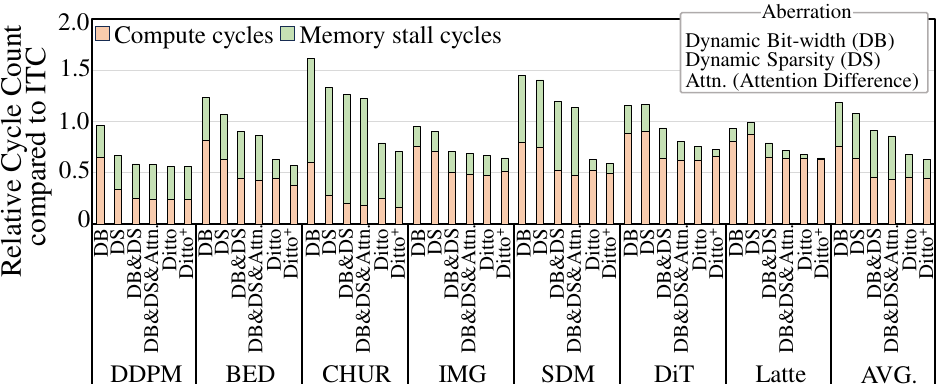}
    \caption{Cycle count results for variation of the Ditto hardware. All designs utilize the layer dependency check technique.}
    \label{fig:eval_ditto}
\end{figure}

Several design space explorations are conducted to analyze the effectiveness of the Ditto algorithm and hardware. 
Firstly, we examine various design choices in Ditto to identify the contribution of each technique, as shown in Fig. \ref{fig:eval_ditto}.
The accelerators that benefit from the Ditto algorithm can be categorized into two types.
One type of accelerator leverages dynamic sparsity (defined as DS) like sparse accelerator~\cite{sparse1,sparse2,spatten}. 
Others utilize dynamic bit-width (defined as DB) like Bit Fusion~\cite{bitfusion} or DRQ~\cite{drq}.
As DS does not support dynamic bit-width, they require multipliers that support the multiplication of 8-bit data and weights. 
Besides, DB is designed to multiply 4-bit data with weight, since DB supports dynamic bit-width.
Then, we categorize the software technique of the Ditto algorithm, as attention difference, Ditto (with Defo), and Ditto$^{+}$ (with Defo$^{+}$).

In the figure, DB and DS exhibit more cycle counts than the baseline ITC.
While DS and DB can improve their performance in computation cycles, they involve high memory stall cycles due to the temporal difference processing.
Combining DB and DS, and applying attention differences can reserve performance improvement over the baseline, but they also suffer from high memory stall cycles, limiting their advantages.
Besides, Ditto and Ditto$^{+}$ can effectively decrease memory stall cycles by applying Defo.
As a result, Ditto shows slightly higher compute cycles than DB\&DS\&Attn. design, but 39.24\%  lower memory stall cycles, achieving 18.32\% performance improvement.

\begin{figure}[t]
    \centering
    \includegraphics[width=1\linewidth]{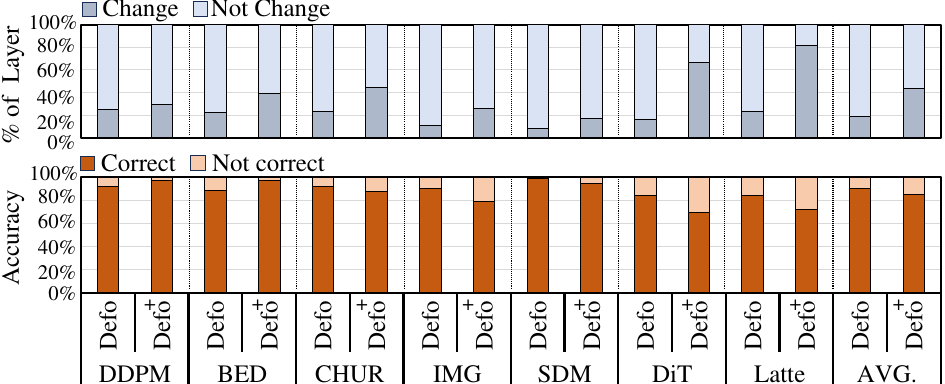}
    \caption{
    Ratio of layer execution types using Defo (top) and accuracy of layer type prediction through Defo (bottom).
    }
    \label{fig:eval_correct}
\end{figure}

To assess the impact of Defo on changes in execution type, a further analysis of Defo is conducted as shown in Fig. \ref{fig:eval_correct}.
The top part of the figure shows the portion of layers that are changed back to their original activation execution through Defo.
In the default Defo, it changes 14.4\% of the layer into the original activation execution on average. Besides, it changes 38.29\% of the layer in Defo$^{+}$.  
Since performance improvements occur in the first time steps by utilizing spatial differences, the threshold to change the execution flow of each layer is reduced.
In particular, 81.6\% of the layer change execution flows into spatial difference processing in the Latte. Since Latte is a video generation task, there is spatial similarity between frames, which means that using spatial differences techniques can be beneficial. 
However, other models show that 63\% of the layers do not change their execution flow even with the Defo$^{+}$. As a result, temporal difference processing is essential to accelerate various diffusion models. 

\begin{figure}[t]
    \centering
    \includegraphics[width=1\linewidth]{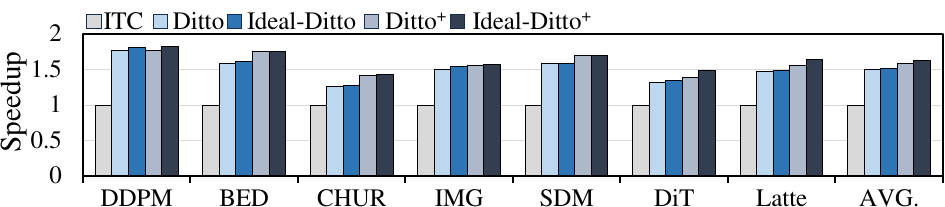}
    \caption{Comparison results of the Ditto and Ditto$^{+}$ hardware with ideal hardware.}
    \label{fig:eval_ideal_dag}
\end{figure}

To evaluate the feasibility of the Defo, we also measure the accuracy of Defo at the bottom of the figure.
The accuracy is measured by assessing whether the execution flow determined by Defo for each layer is indeed the optimal execution flow.
In the figure, Defo and Defo$^{+}$ achieve high accuracy of 92\% and 88.11\%, even if the execution flow is fixed the execution flow at the second time step.
Moreover, we observe that layers where Defo incorrectly determines the optimal execution flow result in only minimal performance degradation, as these layers are on the borderline of the threshold.
To demonstrate the observation, we compare the ideal design with our hardware as shown in Fig. \ref{fig:eval_ideal_dag}. 
In the figure, Ideal-Ditto and ideal-Ditto$^{+}$ represent designs that Defo and Defo$^{+}$ have 100\% accuracy. Therefore, the ideal design always determines the optimal execution flow of each layer for entire time steps.
While the Defo mechanism determines the execution flow in the second time step, Ditto and Ditto$^{+}$ obtains 98.8\% and 95.8\% performance of the ideal design, indicating the feasibility and effectiveness.

\begin{figure}[t]
    \centering
    \includegraphics[width=1\linewidth]{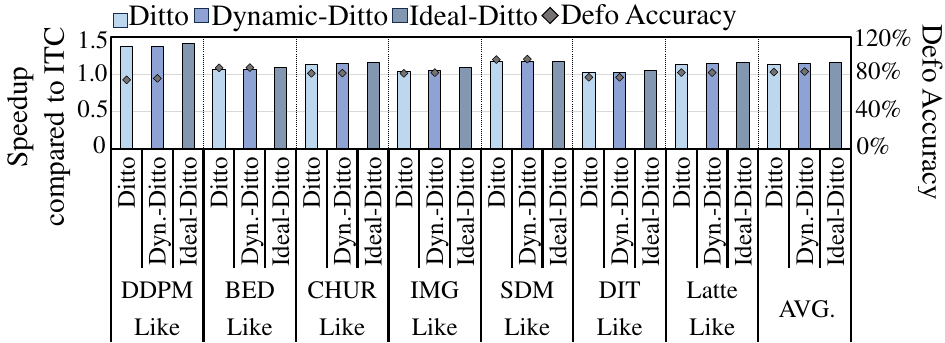}
    \caption{Design space exploration of Defo. Dyn.-Ditto indicate Dynamic Ditto hardware.}
    \label{fig:eval_defo}
\end{figure}

Through our analyses on Fig. \ref{fig:speedup_layer}, we find out that temporal difference processing leads to consistent BOPs reduction compared to original activation processing across all time steps in diffusion models.
However, some future models with high temporal similarity may exhibit dynamic temporal similarity across the time domain, causing BOPs reduction to vary dynamically.
To explore how Ditto can support the case, we adjust the value distribution of our benchmark to make the execution type threshold dynamic and examine the impact.
Moreover, we additionally design the modified Defo algorithm, dynamically determining the execution type of each layer (referred to as Dynamic-Ditto).
We design the Dynamic-Ditto to change the type of processing only from difference processing to original activation processing, since it cannot get the cycle of difference processing during execution with original activation.

In Fig. \ref{fig:eval_defo}, the accuracy of Defo shows a slight decline of 7\% compared to the original benchmark shown in Fig. \ref{fig:eval_correct}.
This is because both Ditto and Dynamic-Ditto predict the benefits of difference processing based on a past time step, while the cycle reduction in the current time step changes dynamically. 
However, Ditto and Dynamic-Ditto achieve 98.03\% and 98.18\% performance of the ideal-Ditto, as the speedup in Defo primarily comes from addressing a few important layers with high memory overhead, which are relatively easy to predict.
Additionally, Dynamic-Ditto slightly outperforms the original Ditto due to the dynamic Defo algorithm, which better adapts to varying conditions. 

\section{Related Work}\label{sec:relatedwork}
Previous works, such as Cambricon-D~\cite{camd} and Guo \textit{et al.}~\cite{isscc}, have explored the temporal similarity between adjacent time steps in diffusion models and leveraged this similarity to reduce the bit-width of layer operations.
However, they exhibit limitation compared to Ditto, since Ditto analyzes and exploits more extensive characteristics of diffusion models.

First, Ditto analyzes temporal differences between the time steps in detail and find out that temporal differences exhibits not only reduced bit-width but also zero values.
By exploiting both dynamic bit-width and sparsity with an lightweight logic, Ditto achieves a higher performance than previous works, which only leverage reduced bit-width. 
While there are other hardware designs that target dynamic sparsity~\cite{spatten,sparse2,sparse1,scnn,s2ta,eureka} or bit-width~\cite{drq,sibia,samsung,bitblade,bitfusion}, they typically focus on one of these technique and do not utilize temporal similarity.

Second, while Cambricon-D~\cite{camd} also addresses the memory overhead of temporal difference processing by proposing a sign-mask data flow approach for specific non-linear functions, Defo dynamically selects the execution type for each layer, effectively managing the memory overhead of the Ditto algorithm. 
Our analysis reveals a consistency in temporal differences across time steps, enabling Defo to adaptively change the execution type for only memory-intensive layers, leaving others at second time step by utilizing lightweight logic. 
Since Defo is independent of the type of non-linear function, it offers greater flexibility compared to prior methods.

Additionally, we introduce Ditto$^{+}$, an enhanced version of Ditto that further exploits spatial similarity in the original activation execution to provide additional performance improvements. 
Ditto$^{+}$ benefits from the Ditto hardware, which is not limited to exploiting sparsity and bit-width reductions only in temporal differences.
While there are previous works that explore either spatial~\cite{diffy,pointcloud,reuse1,slid,deepreuse} or temporal similarity~\cite{edgedrnn,deltarnn,camd,fuzzy,video,isscc}, they generally support only one of these aspects. 
In contrast, our hardware and execution flow optimization allow for the simultaneous exploitation of both spatial and temporal similarities, offering a significant performance advantage.

Lastly, previous works like Q-Diffusion~\cite{qdiff} and TDQ~\cite{tdq} have introduced timestep-specific quantization methods for diffusion models by leveraging the varying value distribution of input activations across time steps.
Ditto enable synergy with them by combining quantization with the exploitation of temporal similarities.
Integrating our scheme with existing diffusion model quantization methods~\cite{qdiff,tdq,ptqd,ptq4dm,tfmq} allows Ditto to further accelerate quantized diffusion models, significantly reducing denoising latency of diffusion models.
\section{conclusion}
Diffusion models are state-of-the-art \ac{DNN} algorithms for image generation but suffer from long execution times and high computational overhead due to recursive time steps. 
This paper introduces high temporal value similarity between adjacent time steps and reveals that the high similarity exhibits a narrower value range of differences between the time steps.
We observe that the smaller value range exposes reduced bit-width and zero values in the diffusion models with quantization.
Based on our observations, we propose the Ditto algorithm to reduce computation overhead by utilizing zero-skipping and reduced bit-width. 
Since the temporal difference processing incurs memory overhead in some layers, the algorithm is further optimized by Defo, automatically determining the optimal execution flow of each layer.
We also design the Ditto hardware that supports dynamic bit-width, sparsity, and adaptive execution flow, achieving up to 1.5$\times$ speedup and 17.74\% energy saving over the baseline hardware.

\section*{Acknowledgment}
This work was supported by the Institute of Information \& Communications Technology Planning \& Evaluation (IITP), supported by the Korean government (Ministry of Science and ICT, MSIT) (No. 2024-0-00441, Memory-Centric Architecture Using Reconfigurable PIM Devices) and by Samsung Electronics Co., Ltd.
Won Woo Ro is the corresponding author.

\bibliographystyle{IEEEtranS}
\bibliography{refs}

\end{document}